\documentclass[aps,pra,amsmath,twocolumn,amssymb,superscriptaddress]{revtex4-1}

\usepackage{amssymb,amsfonts,amsmath}

\usepackage{color}
\usepackage[apple mac]{inputenc}
\usepackage[english]{babel}
\usepackage{times}
\usepackage{latexsym}
\usepackage{fancyhdr}
\usepackage{verbatim}
\usepackage{graphicx}
\usepackage{epsf}
\usepackage{subfigure}
\usepackage{bm}
\usepackage{epstopdf}\usepackage[
top    = 1.3cm,
bottom = 1.2cm,
left   = 1.5cm,
right  = 1.3cm]{geometry}

\definecolor{Nathanblue}{rgb}{0.,0.24,0.51}
\newcommand{\blue}{\color{Nathanblue}}

\definecolor{orange}{rgb}{0.96,0.24,0.00}

\def\be{\begin{equation}}
\def\ee{\end{equation}}
\def\bs#1{\boldsymbol{#1}}

\begin{document}

\title{ {\blue Probing topology by ``heating": \\ Quantized circular dichroism in ultracold atoms}}

\author{D.~T. Tran}
\affiliation{Center for Nonlinear Phenomena and Complex Systems,
Universit\'e Libre de Bruxelles, CP 231, Campus Plaine, B-1050 Brussels, Belgium}

\author{A. Dauphin}
\affiliation{ICFO-Institut de Ciencies Fotoniques, The Barcelona Institute of Science and Technology, 08860 Castelldefels (Barcelona), Spain}

\author{A. G. Grushin}
\affiliation{Department of Physics, University of California, Berkeley, California 94720, USA}
\affiliation{Institut N\'eel, CNRS and Université Grenoble Alpes, F-38042 Grenoble, France}


\author{P. Zoller}
\affiliation{International Solvay Institutes,
Universit\'e Libre de Bruxelles, Campus Plaine, B-1050 Brussels, Belgium}
\affiliation{Institute for Theoretical Physics, University of Innsbruck, A-6020 Innsbruck, Austria}
\affiliation{Institute for Quantum Optics and Quantum Information of the Austrian Academy of Sciences, A-6020 Innsbruck, Austria}

\author{N. Goldman${\blue ^*}$}
\affiliation{Center for Nonlinear Phenomena and Complex Systems,
Universit\'e Libre de Bruxelles, CP 231, Campus Plaine, B-1050 Brussels, Belgium}

\date{\today}

\begin{abstract}

We reveal an intriguing manifestation of topology, which appears in the depletion rate of topological states of matter in response to an external drive. This phenomenon is presented by analyzing the response of a generic 2D Chern insulator subjected to a circular time-periodic perturbation:~due to the system's chiral nature, the depletion rate is shown to depend on the orientation of the circular shake. Most importantly, taking the difference between the rates obtained from two opposite orientations of the drive, and integrating over a proper drive-frequency range, provides a direct measure of the topological Chern number of the populated band ($\nu$):~this ``differential integrated rate" is directly related to the strength of the driving field through the quantized coefficient $\eta_0\!=\!\nu /\hbar^2$. Contrary to the integer quantum Hall effect, this quantized response is found to be non-linear with respect to the strength of the driving field and it explicitly involves inter-band transitions. We investigate the possibility of probing this phenomenon in ultracold gases and highlight the crucial role played by edge states in this effect. We extend our results to 3D lattices, establishing a link between depletion rates and the non-linear photogalvanic effect predicted for Weyl semimetals. The quantized circular dichroism revealed in this work designates depletion-rate measurements as a universal probe for topological order in quantum matter.

\end{abstract}

\maketitle

The quantization of physical observables plays a central role in our understanding and appreciation of Nature's laws, as was already evidenced by the antique work of Pythagoras on harmonic series, and many centuries later, by the identification of the Balmer series in atomic physics~\cite{foot}. More recently, in condensed-matter physics, the observation of quantized conductance unambiguously demonstrated the quantum nature of matter, in particular, the possibility for electronic currents to flow according to a finite set of conducting channels~\cite{Klitzing,Beenakker}. While the quantized plateaus depicted by the conductance of mesoscopic channels depend on the samples geometry~\cite{Beenakker}, a more universal behavior exists when a two-dimensional electron gas is immersed in an intense magnetic field~\cite{Klitzing}:~In the non-interacting regime, the Hall conductivity is then quantized according to the Thouless-Kohmoto-Nightingale-Nijs (TKNN) formula~\cite{TKNN}, $\sigma_{\text{H}}\!=\!(e^2/h) \nu$, where $h$ is Planck's constant and where $\nu$ is a topological invariant --  the Chern number -- associated with the filled Bloch bands~\cite{Niu1985,Kohmoto}. Since the discovery of this integer quantum Hall (QH) effect, the intimate connection between topology and quantized responses has been widely explored in solid-state physics~\cite{topological_review,Xiao_review}, revealing remarkable effects such as the quantization of Faraday rotation in 3D topological insulators~\cite{quantized_faraday}.

Building on their universal nature, topological properties are currently studied in an even broader context~\cite{Asorey}, ranging from ultracold atomic gases~\cite{Goldman_progress} and photonics~\cite{photon_topological2,photon_topological} to mechanical systems~\cite{Huber_review}. Interestingly, these complementary and versatile platforms offer the possibility of revealing unique topological properties, such as those emanating from engineered dissipation~\cite{Rudner_dissipation,Diehl_NatPhys,Alberti_dissipation}, time-periodic modulations~\cite{Kitagawa_class,Floquet_photonics,Genuine_Floquet_exp,JotzuNat,AidelsburgerNat,Flaschner2016,Lindner2017}, quantum walks~\cite{Kitagawa_walks,Walks_Dauphin_exp} and controllable interactions~\cite{Goldman_progress,topology_interactions,topology_interactions_Cooper}. In ultracold gases, the equivalent of the TKNN formula was explored by visualizing the transverse displacement of an atomic cloud in response to an applied force~\cite{AidelsburgerNat}; the Chern number $\nu$, and the underlying Berry curvature~\cite{Xiao_review}, were also extracted through state-tomography~\cite{Flaschner,4D_Spielman}, interferometry~\cite{Duca}, and spin-polarisation measurements~\cite{Wu_science}. Besides, the propagation of robust chiral edge modes was identified in a variety of physical platforms~\cite{Asorey,Goldman_progress, photon_topological2,photon_topological,Huber_review}.

\begin{figure}[h!]
\resizebox{0.48\textwidth}{!}{\includegraphics*{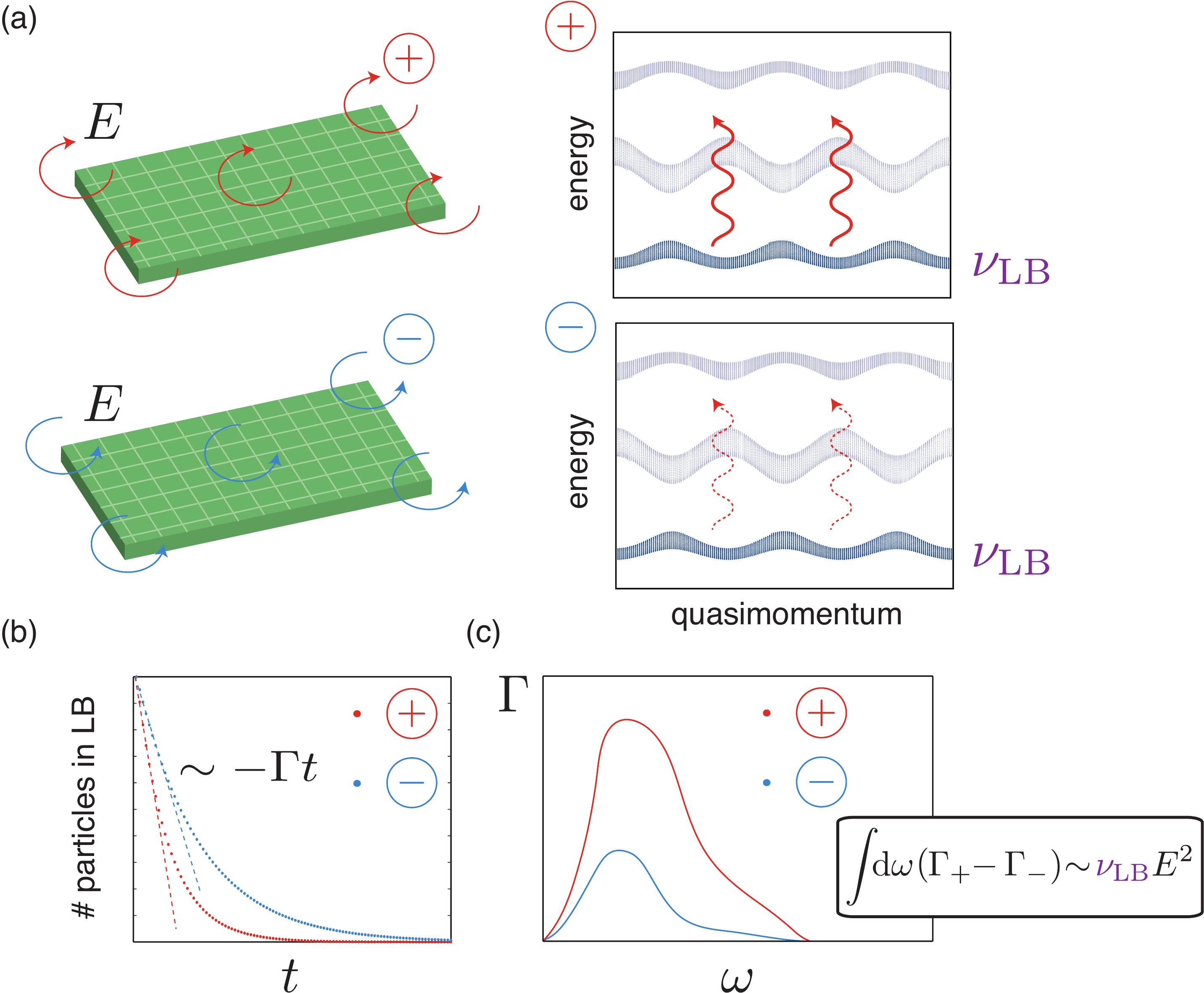}}  
\caption{Topology through heating:~(a) A 2D Fermi gas is initially prepared in the lowest Bloch band (LB) of a lattice, with Chern number $\nu_{\text{LB}}$, and it is then subjected to a circular time-periodic modulation  [Eq.~\eqref{Htime}]. (b) The rate $\Gamma$ associated with the depletion of the populated band, $\delta N_{\text{LB}}(t)\!\approx\!-\Gamma t$, is found to depend on the orientation of the drive, $\Gamma_+\!\ne\!\Gamma_-$, whenever the LB is characterized by a non-trivial Chern number $\nu_{\text{LB}}\!\ne\!0$. (c) Integrating the differential rate over a relevant drive-frequency range, $\Delta \Gamma^{\text{int}}\!=\!\int \text{d} \omega \left ( \Gamma_+\!-\!\Gamma_-\right )/2$,  leads to a quantized result, $\Delta \Gamma^{\text{int}}/A_{\text{syst}}\!=\!(\nu_{\text{LB}}/\hbar^2) E^2 $, where $E$ is the strength of the drive and $A_{\text{syst}}$ is the system's area [Eq.~\eqref{main_result}]. }
 \label{fig:1} 
\end{figure} 

In this work, we demonstrate that the depletion rate of a Bloch band in a quantum lattice system, which reflects the inter-band (dissipative) response to a time-dependent perturbation, satisfies a quantization law imposed by topological properties. This observation of depletion-rate quantization suggests that \emph{heating a system can be exploited to extract its topological order}. Specifically, our method builds on the chiral nature of systems featuring Bloch bands with non-zero Chern number~\cite{topological_review}.~First, we find that the depletion rate of a circularly-shaken Chern insulator, as captured by Fermi's Golden Rule (FGR), crucially depends on the orientation (chirality) of the drive~[Fig.~\ref{fig:1}]. Then, we identify an intriguing quantization law for the \emph{differential integrated rate} (DIR), $\Delta \Gamma^{\text{int}}$, which is defined as the difference between the rates obtained from opposite orientations of the drive, integrated over a relevant frequency range. The quantization of the DIR can be simply expressed as
\be
\Delta \Gamma^{\text{int}}/A_{\text{syst}} = \eta_0 E^2,  \qquad   \eta_0=(1/\hbar^2) \,  \nu ,\label{main_result}
\ee
in terms of the drive amplitude $E$ and the topological response coefficient $\eta_0$; here $\nu$ denotes the Chern number of the populated band and $A_{\text{syst}}$ is the system's area. This result is in agreement with the intuition that the response of a trivial insulator ($\nu\!=\!0$) to a circular drive should not depend on the latter's orientation. Interestingly, the quantized response~\eqref{main_result} identified in this work is non-linear with respect to the strength of the driving field $E$, and it explicitly involves inter-band transitions~\cite{Culcer}, indicating that this phenomenon is distinct in essence from the TKNN paradigm~\cite{TKNN}, which is associated with linear transport and captured by single-band semiclassics~\cite{Xiao_review,PZO16}.  This quantized effect related to circular dichroism establishes depletion-rate measurements as a versatile probe for topological order.

In the following, we demonstrate how the depletion rate associated with circularly-driven Chern insulators can be related to the topological Chern number $\nu$, and we identify the relevant response coefficient in this context, $\eta_0\!=\!\nu/\hbar^2$. We explain how this result relates to the general concept of circular dichroism, through the universal Kramers-Kronig relations~\cite{Jackson}. We then discuss how this effect could be probed in realistic cold-atom systems, setting the focus on how to avoid the detrimental contribution of edge-states in this framework. We then extend our results to 3D lattices, providing an instructive connection between topological depletion rates and the non-linear photogalvanic effect~\cite{Sipe} recently predicted for Weyl semimetals~\cite{JGM11}. Finally, concluding remarks and perspectives are presented.

\section*{Results}

\subsection*{Topology and quantization of integrated depletion rates}

We start by studying a non-interacting (spinless) gas in a generic two-dimensional (2D) lattice, as described by the single-particle Hamiltonian $\hat H_0$. In our study, we will assume that the lowest Bloch band (LB) of the spectrum, which is separated from higher bands by a bulk gap $\Delta_{\text{gap}}$, is initially completely filled with fermions; the following discussion can be straightforwardly extended to other initial filling conditions. Considering systems with broken time-reversal symmetry, the topological properties of this LB will be accurately captured by the Chern number~\cite{Niu1985,Kohmoto,topological_review}, henceforth denoted $\nu_{\text{LB}}$. Thus, subjecting this system to a constant electric field $\bs E\!=\!E_y \bs{1}_y$ generates a total Hall current $\bs J\!=\!J^x \bs{1}_x$ satisfying the TKNN formula~\cite{TKNN}:
\be
J^x/A_{\text{syst}}=j^x=\sigma_{\text{H}} E_y, \qquad \sigma_{\text{H}}=(q^2/h) \, \nu_{\text{LB}},\label{current}
\ee
where $A_{\text{syst}}$ is the system's area and $q$ is the charge of the carriers ($q\!=\!e$ in an electron gas). In gases of neutral atoms, such transport equations can be probed by measuring the flow of particles~\cite{Brantut,Munich_current,Catherine_current,PZO16} in response to a synthetic electric field (e.g.~an optical gradient~\cite{AidelsburgerNat}); this latter situation corresponds to setting $q\!=\!1$ in Eq.~\eqref{current}.

In this work, however, we are interested in the \emph{depletion rate} of this system in response to a circular time-periodic perturbation, as described by the total time-dependent Hamiltonian
\begin{align}
&\hat H_{\pm} (t)\!=\! \hat H_0 \!+\!2 E \, \left [\cos (\omega t) \hat x \pm \sin (\omega t) \hat y \right ] , \label{Htime}
\end{align}
where $\pm$ refers to the two possible orientations (chirality) of the drive, $(\hat x,\hat y)$ are the position operators~\cite{footnote_position}, and $\omega$ is a frequency used to drive inter-band transitions. Such circular shaking of 2D lattices can be implemented in cold atoms trapped in optical lattices~\cite{Eckardt}, using piezo-electric actuators~\cite{JotzuNat};~the Hamiltonian in Eq.~\eqref{Htime} equally describes electronic systems subjected to circularly-polarized light~\cite{Bennett,Aoki,Lindner,Song,JGM11}.~The total number of particles scattered and extracted from the LB, $N_{\pm}(\omega,t)\!\approx\!\Gamma_{\pm} (\omega) t$, is associated with the depletion rate $\Gamma_{\pm}$ [Fig.~\ref{fig:1}], which can be accurately evaluated using Fermi's Golden Rule~\cite{CCT_book}
\be
\Gamma_{\pm} (\omega)= \frac{2 \pi}{\hbar} E^2 \sum_{e \notin \text{LB}} \sum_{g \in \text{LB}} \vert \langle e \vert \hat x \pm i \hat y \vert g \rangle \vert^2 \delta^{(t)} (\varepsilon_e - \varepsilon_g -  \hbar \omega) , \label{rate_gen_real}
\ee
where  $\vert g \rangle$ [resp.~$\vert e \rangle$] denotes all the initially occupied [resp.~unoccupied] single-particle states with energy $\varepsilon_{g}$ [resp.~$\varepsilon_{e}$], 	and  where $\delta^{(t)}(\varepsilon)\!=\!(2\hbar/\pi t) \sin^2 (\varepsilon t/2\hbar)/\varepsilon^2\!\longrightarrow\!\delta(\varepsilon)$ in the long-time limit~\cite{footnote:Rabi}. The transitions to initially occupied states ($g$) are excluded in Eq.~\eqref{rate_gen_real}, as required by Fermi statistics, which is an important feature when many bands are initially occupied.  We also point out that the number of scattered particles $N_{\pm}(\omega,t)$ can be directly detected in cold atoms by measuring the dynamical repopulation of the bands through band-mapping techniques, as was demonstrated for non-trivial Chern bands in Ref.~\cite{AidelsburgerNat}. 

For the sake of pedagogy, let us first analyze the excitation rate~\eqref{rate_gen_real} in a frame where the total Hamiltonian~\eqref{Htime} is translationally invariant. Performing the frame transformation generated by the operator
\be
\hat R_{\pm}=\exp \left \{i  \frac{2 E}{\hbar \omega} \left [ \sin (\omega t) \hat x \mp \cos (\omega t) \hat y \right ] \right \},
\ee
the time-dependent Hamiltonian~\eqref{Htime} is modified according to
\begin{align}
&\hat{\mathcal{H}}_{\pm} (t)\!\approx\! \hat H_0(\bs k) \!+\! \frac{2E}{\hbar \omega}\! \left \{ \sin (\omega t) \frac{\partial \hat H_0 (\bs k)}{\partial k_x} \!\mp\! \cos (\omega t ) \frac{\partial \hat H_0 (\bs k)}{\partial k_y}\!\right \},\label{Htrans}
\end{align}
where we now adopted the momentum representation, and omitted higher-order terms in $E$ in agreement with the perturbative approach~\cite{CCT_book} inherent to the FGR in Eq.~\eqref{rate_gen_real} and below. In this frame, the depletion rate~\eqref{rate_gen_real} now takes the more suggestive form
\begin{align}
&\Gamma_{\pm} (\omega)=\sum_{\bs k}\Gamma_{\pm} (\bs k; \omega), \label{rate_gen} \\
&\Gamma_{\pm} (\bs k; \omega)\!=\! \frac{2 \pi}{\hbar}\sum_{n > 0} \, \vert \mathcal{V}_{n0}^\pm(\bs k)\vert ^2 \, \delta^{(t)} (\varepsilon_n(\bs k) \!-\! \varepsilon_0 (\bs k) \!-\!  \hbar\omega), \notag \\
&\vert \mathcal{V}_{n0}^\pm(\bs k)\vert ^2=\left (E/\hbar \omega \right )^2 \bigg \vert \bigg \langle n (\bs k) \bigg \vert  \frac{1}{i} \frac{\partial \hat H_0}{\partial k_x}\!\mp\! \frac{\partial \hat H_0}{\partial k_y} \bigg \vert 0 (\bs k)   \bigg \rangle \bigg \vert^2.\notag
\end{align}
Here, we introduced the initially-populated Bloch states of the LB, $\vert g\rangle\!\equiv\!\vert 0 (\bs k) \rangle$, of dispersion $\varepsilon_0(\bs k)$, as well as the initially-unoccupied Bloch states of the higher bands, $\vert e\rangle\!\equiv\!\vert n (\bs k) \rangle$, of dispersion $\varepsilon_n(\bs k)$ and band index $n$. We note that in an ideal translationally-invariant non-interacting system, the inter-band transitions occurring at each ${\bs k}$ yield Rabi oscillations~\cite{CCT_book}, hence leading to a linear growth of the depletion rates $\Gamma_{\pm} (\bs k; \omega)$; we point out that this effect, which is naturally damped in solid-state systems through disorder~\cite{JGM11}, is in fact irrelevant when integrating the depletion rates over the drive frequency, as we now discuss.

Integrating the depletion rates $\Gamma_{\pm} (\omega)$ in Eq.~\eqref{rate_gen} over all drive frequencies $\omega\!\ge\! \Delta_{\text{gap}}/\hbar$, i.e.~activating all possible transitions between the filled LB and the higher bands~\cite{footnote_omega_range},  and considering the difference between these integrated rates, $\Delta \Gamma^{\text{int}}=(\Gamma^{\text{int}}_+-\Gamma^{\text{int}}_-)/2$, defines the \emph{differential integrated rate} (DIR), which reads
\begin{align}
\Delta \Gamma^{\text{int}}&=4 \pi (E/\hbar)^2 \,  \text{Im} \!  \sum_{n > 0} \sum_{\bs k}\, \frac{  \langle 0 \vert \partial_{k_x}\hat H_0 \vert n \rangle \langle n \vert \partial_{k_y}\hat H_0 \vert 0 \rangle }{(\varepsilon_0 - \varepsilon_n)^2}.\label{general}
\end{align}
Comparing the latter with the expression for the Chern number~\cite{Xiao_review}
\begin{align}
\nu_{\text{LB}}&=  \frac{4 \pi}{A_{\text{syst}}} \text{Im}   \sum_{n > 0} \sum_{\bs k} \frac{  \langle 0 \vert \partial_{k_x}\hat H_0 \vert n \rangle \langle n \vert \partial_{k_y}\hat H_0 \vert 0 \rangle }{(\varepsilon_0 - \varepsilon_n)^2},\label{chern}
\end{align}
we obtain the simple quantization law for the DIR per unit area in Eq.~\eqref{main_result}, with $\nu\!=\!\nu_{\text{LB}}$.

Remarkably, the integration inherent to the definition of the DIR reveals the Chern number of the ground band, while the properties of excited states drop out through the summation over all final states. The relation in Eq.~\eqref{main_result} is reminiscent of the transport equation~\eqref{current}  associated with the QH effect:~the DIR per unit area is directly related to the driving field $E$ through a response coefficient $\eta_0$ that only depends on the topology of the populated band and on a universal constant ($\hbar^{-2}$). We point out that, contrary to the linear transport equation~\eqref{current}, the quantized response in Eq.~\eqref{main_result} is \emph{non-linear} with respect to the driving field, which highlights its distinct origin. In particular, the differential aspect of the measurement, which directly probes the chirality of the system by comparing its response to opposite shaking orientations, plays an essential role in this distinct quantized effect. Besides, we note that the latter explicitly involves inter-band transitions, ruling out the possibility of capturing it through a single-band semiclassical approach~\cite{Xiao_review}. It is straightforward to generalize the result in Eq.~\eqref{main_result} to situations where many bands are initially populated, in which case $\nu_{\text{LB}}$ should be replaced by the sum over the Chern numbers associated with these bands.

In the case of two-band models ($n\!=\!1$), we point out that the \emph{local} differential rate $\Delta\Gamma (\bs k; \omega)\!=\![\Gamma_+ (\bs k; \omega)\!-\!\Gamma_- (\bs k; \omega)]/2$ resulting from Eq.~\eqref{rate_gen} is directly proportional to the Berry curvature $\Omega(\bs k)$ of the LB~\cite{Xiao_review}. Hence, measuring $\Delta\Gamma (\bs k; \omega)$ from wave packets (prepared in the LB and centered around $\bs k$) offers an elegant method to directly probe the geometrical properties of Bloch bands, as captured by the local Berry curvature~\cite{Xiao_review}. Also, in that case, the allowed transitions are automatically restricted to $\vert 0 (\bs k) \rangle\!\rightarrow\vert 1 (\bs k) \rangle$, irrespective of Fermi statistics.

Practically, we propose that the integrated rates could be experimentally extracted from many individual depletion-rate-measurements~\cite{AidelsburgerNat,footnote_omega_range}, corresponding to sampled (fixed) values of $\omega$:~$\Gamma_{\pm}^{\text{int}}\!\approx\!\sum_l \Gamma_{\pm}(\omega_l) \Delta_{\omega}$.~This scheme could also be facilitated by the use of multi-frequency drives; see Ref.~\cite{Schuler} for a very recent application of our scheme based on short pulses.

\begin{figure}[h!]
\resizebox{0.48\textwidth}{!}{\includegraphics*{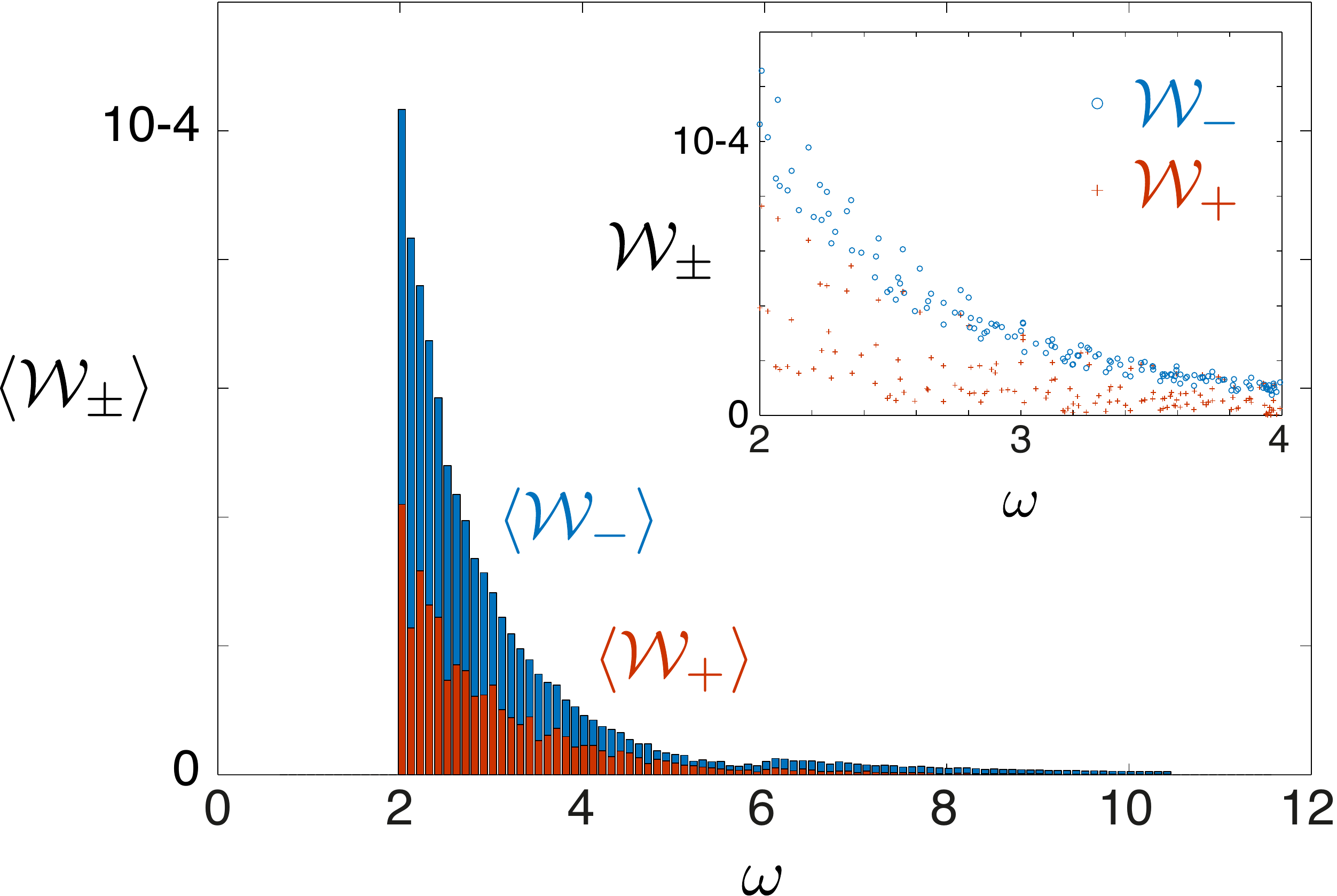}}  
\caption{Transition matrix elements for the driven two-band Haldane model with $10^{4}$ lattice sites and PBC.  Specifically, the inset shows the  matrix elements $\mathcal{W}_{\pm}\!=\!(2 \pi/\hbar) \vert \mathcal{V}_{10}^\pm (\bs k) \vert ^2$, as defined in Eq.~\eqref{rate_gen}, for all possible transitions, $\omega\!=\!\left [\varepsilon_1(\bs k)\!-\!\varepsilon_0 (\bs k)\right ]/\hbar$; the main plot shows the averaged values $\langle \mathcal{W}_{\pm} (\omega) \rangle$, defined within each interval of width $\Delta_{\omega}\!=\!0.1J/\hbar$. The model parameters are set such that $\Delta_{\text{gap}}\!\approx\!2J$, where $J$ is the nearest-neighbor hopping amplitude; the strength of the drive [Eq.~\eqref{Htime}] is $E\!=\!0.001 J/d$, where $d$ is the lattice spacing. The DIR [Eq.~\eqref{general}] obtained from this numerical data  yields $\Delta \Gamma^{\text{int}}(\hbar^2/A_{\text{syst}}E^2)\!\approx\!-1.00$, in agreement with Eq.~\eqref{main_result} and the theoretical prediction $\nu_{\text{LB}}\!=\!-1$.  The matrix elements $\mathcal{W}_{\pm}$ are expressed in units of $J^2/\hbar$, while the frequency  $\omega$ is given in units of $J/\hbar$.
}
 \label{fig:2} 
\end{figure}

We have validated the quantization law in Eq.~\eqref{main_result}, based on a numerical study of the two-band Haldane model~\cite{Haldane}, in the topological phase where $\nu_{\text{LB}}\!=\!-1$. The matrix elements $\mathcal{W}_{\pm}\!=\!(2 \pi/\hbar) \vert \mathcal{V}_{10}^\pm (\bs k) \vert ^2$, as defined in Eq.~\eqref{rate_gen}, were calculated for a honeycomb lattice of size $100\!\times\!100$, with periodic boundary conditions (PBC); see Fig.~\ref{fig:2}. We verified that the DIR [Eq.~\eqref{general}], as evaluated from this numerical data and from the density of states, yields $\Delta \Gamma^{\text{int}}(\hbar^2/A_{\text{syst}}E^2)\!\approx\!-1.00$, in perfect agreement with the quantized prediction of Eq.~\eqref{main_result}; see also the Supplementary Materials.

\subsection*{Relation to circular dichroism and Kramers-Kronig relations}

Interestingly, the result in Eq.~\eqref{main_result} is deeply connected to the well-known Kramers-Kronig relations~\cite{Jackson}, which are a direct consequence of the causal nature of response functions~\cite{Bennett,Hu}. Considering the conductivity tensor $\sigma^{a b}$, the Kramers-Kronig relations take the form~\cite{Bennett}
 \begin{align}
&\sigma^{a b}_R (\omega)= (2/\pi)  \int_{0}^{\infty} \,  \frac{\tilde \omega \, \sigma^{a b}_I (\tilde \omega)}{\tilde{\omega}^2 - \omega^2} \, \text{d} \tilde \omega , \label{Kramers} 
 \end{align}
where $\sigma^{a b}\!=\! \sigma^{a b}_R+i \sigma^{a b}_I$ has been separated into real and imaginary parts, and where $a,b\!=\!(x,y)$. In the limit  $\omega \rightarrow 0$, the relation~\eqref{Kramers} yields the sum rule 
\begin{equation}
\sigma_{\text{H}} =\lim_{\omega\rightarrow0}\sigma^{x y}_R (\omega)=(2/\pi)  \int_{0}^{\infty} \,   \tilde{\omega}^{-1} \, \sigma^{x y}_I (\tilde \omega) \, \text{d} \tilde \omega.\label{sum_rule}
\end{equation} 
Besides, following Bennett and Stern~\cite{Bennett}, the power absorbed by a system subjected to the circular time-dependent perturbation in Eq.~\eqref{Htime} can be related to the conductivity tensor as 
\begin{equation}
P_{\pm}(\omega)\!=\! 4 A_{\text{syst}} E^2 \left [ \sigma^{x x}_R(\omega) \pm \sigma^{x y}_I (\omega) \right ], \label{power_eq}
\end{equation}
where $\pm$ again refers to the orientation of the drive. Relating the depletion rate to the absorbed power, $\Gamma_{\pm} (\omega)\!=\! P_{\pm}(\omega)/\hbar \omega$, and introducing the differential rate $\Delta \Gamma=(\Gamma_+-\Gamma_-)/2$, then yields the useful relation
\begin{equation}
\sigma^{x y}_I (\omega)=\hbar \omega \Delta \Gamma (\omega) / 4 A_{\text{syst}} E^2 .\label{diff_rate_eq}
\end{equation}
Finally, inserting Eq.~\eqref{diff_rate_eq} into \eqref{sum_rule} allows one to directly relate the DIR to the Hall conductivity of the probed system:
\begin{equation}
\Delta \Gamma^{\text{int}}/A_{\text{syst}}=(1/A_{\text{syst}}) \int_{0}^{\infty}  \Delta \Gamma (\omega) \, \text{d} \omega =(2\pi E^2/\hbar) \, \sigma_{\text{H}}  . \label{final_dichroism_rate}
\end{equation}
Importantly, the general expression~\eqref{final_dichroism_rate} leads to the quantization law in Eq.~\eqref{main_result}, when considering the TKNN formula for the Hall conductivity $\sigma_{\text{H}}$ of a Chern insulator  [see Eq.~\eqref{current}]. We note that other intriguing sum rules have been identified in the context of circular dichroism~\cite{Souza}, and that these could be exploited to access useful ground-state properties (e.g.~the orbital magnetization of insulators).


\subsection*{On the effects of boundaries}

Importantly, the derivation leading to Eq.~\eqref{main_result} implicitly assumed translational invariance and PBC (i.e.~a torus geometry);~in particular, this result disregards the effects related to the presence of (chiral) edge states in finite lattices~\cite{topological_review}. Here, we reveal the important contribution of edge states, when considering more realistic systems with boundaries.

In order to analyze lattices with edges (and more generally, systems that do not present translational symmetry, such as disordered systems~\cite{Bianco} or quasicrystals~\cite{Duc_Thanh}), it is instructive to expand the modulus squared in the ``real-space"  formula~\eqref{rate_gen_real} and then to integrate the latter over all frequencies $\omega$;~this yields the integrated rates
\begin{equation}
\Gamma^{\text{int}}_{\pm}= (2 \pi/\hbar^2) E^2 \sum_{g \in \text{LB}} \langle g \vert \hat P (\hat x \mp i \hat y) \hat Q (\hat x \pm i \hat y) \hat P \vert g \rangle ,\label{int_rate_real}
\end{equation}
where we introduced the projector $\hat P\!=\!1\!-\!\hat Q$ onto the LB. Then, the expression for the DIR [Eq.~\eqref{general}] now takes the form
\begin{align}
&\Delta \Gamma^{\text{int}} \!=\!(\Gamma^{\text{int}}_+\!-\!\Gamma^{\text{int}}_-)/2= (E/\hbar)^2 \,  \text{Tr } \hat{\mathfrak{C}}, \label{real_space_result} \\ 
&\hat{\mathfrak{C}}= 4 \pi \, \text{Im} \hat P \hat x \hat Q \hat y \hat P ,\notag
\end{align}
where $\text{Tr}(\cdot)$ is the trace. Importantly, when applying PBC, the quantity $(1/A_{\text{syst}})\text{Tr } \hat{\mathfrak{C}} \!=\! \nu_{\text{LB}}$ is equal to the Chern number of the populated band~\cite{Kitaev,Bianco}, such that the result in Eq.~\eqref{main_result} is indeed recovered in this real-space picture;~in particular, this demonstrates the applicability of Eq.~\eqref{main_result} to systems without translational symmetry and finite observation times.

\begin{figure}[!]
\resizebox{0.48\textwidth}{!}{\includegraphics*{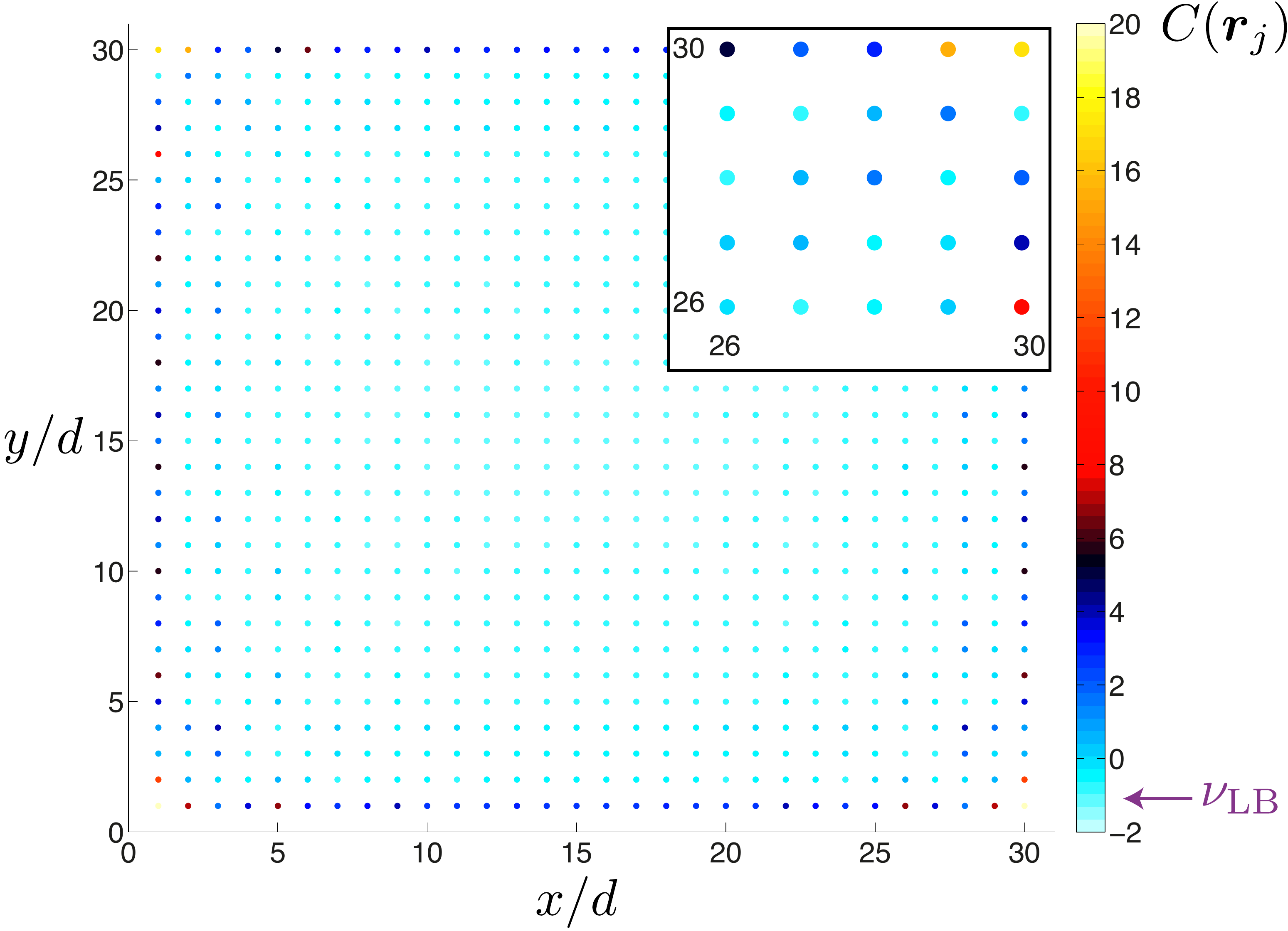}}  
\caption{Local Chern marker $C(\bs r_j)$ in a 2D lattice with boundaries (OBC) realizing the Haldane model; $d$ is the lattice spacing. Far from the boundaries, the marker is $C(\bs r_j)\!\approx\!-1$, in agreement with the Chern number of the populated band $\nu_{\text{LB}}\!=\!-1$. Close to the edges, the local marker is very large and positive (see the zoom shown in the inset) such that the total contribution of the edges exactly cancels the bulk contribution, $ \sum_{\bs r_j} C(\bs r_j) \!=0$:~the DIR in Eq.~\eqref{dir_space_local} vanishes in a system with boundaries.}
 \label{fig:3} 
\end{figure}

The real-space approach allows for the identification of the strong edge-states contribution to the DIR $\Delta \Gamma^{\text{int}}$, when (realistic) open boundary conditions (OBC) are considered. To see this, let us recall that the trace in Eq.~\eqref{real_space_result} can be performed using the position (or Wannier-state) basis $\{\vert \bs r_j \rangle \}$; in particular, inspired by Refs.~\cite{Bianco,Duc_Thanh}, we decompose the DIR~\eqref{real_space_result} in terms of bulk and edge contributions:
\begin{align}
\Delta \Gamma_{\text{OBC}}^{\text{int}}&=(E/\hbar)^2 \left \{ \sum_{\bs r_j \in \text{bulk}} C(\bs r_j) \!+\! \sum_{\bs r_j \in \text{edge}} C(\bs r_j)\right \},\label{dir_space_local}
\end{align}
where we introduced the local marker $C(\bs r_j)\!=\!\langle \bs r_j \vert  \hat{\mathfrak{C}} \vert \bs r_j \rangle$. As illustrated in Fig.~\ref{fig:3}, the local marker $C(\bs r_j)\!\approx\!\nu_{\text{LB}}$ is almost perfectly uniform within the bulk of the system; in the thermodynamic limit, the bulk contribution $\left (\sum_{\bs r_j \in \text{bulk}} C(\bs r_j)\!\rightarrow\!A_{\text{syst}}\nu_{\text{LB}}\right )$ leads to the quantized DIR predicted by Eq.~\eqref{main_result} for PBC. However, the distinct contribution of the edge states, which is clearly identified at the boundaries in Fig.~\ref{fig:3}, is found to exactly compensate the bulk contribution; see also Refs.~\cite{Bianco,Souza}. Consequently, the total DIR in Eq.~\eqref{dir_space_local} vanishes for OBC, $\Delta \Gamma_{\text{OBC}}^{\text{int}}\!=\!0$, which is in agreement with the triviality of the underlying fibre bundle (the corresponding base space being flat~\cite{Nakahara}). This important observation shows the drastic role played by the boundary in the present context; in particular, it indicates that the edge-states contribution must be annihilated in order to observe the quantized DIR [Eq.~\eqref{main_result}] in experiments, as we further investigate in the  paragraphs below. 

Before doing so, let us emphasize that the detrimental contribution of the edge states cannot be simply avoided by performing a local measurement in the bulk, far from the edges. Indeed, probing the DIR in some region $R$ would formally correspond to evaluating the quantity $\mu_{R}\!=\!(1/A_{\text{syst}})\text{Tr } \tilde{\mathfrak{C}}$, where $\tilde{\mathfrak{C}}\!=\! 4 \pi \text{Im}\left ( \hat P \hat R \hat x \hat Q \hat R \hat y  \hat P \right )$, and where $\hat R$ projects onto the region $R$. While the local Chern number~\cite{Bianco}, defined as $\nu_{R}\!=\!(4 \pi/A_{\text{syst}})\text{Tr }  \text{Im} \left (\hat R \hat P  \hat x \hat Q \hat y \hat P\right )\!\approx\!\nu_{\text{LB}}$, can indeed provide an approximate value for the Chern number of the LB, we find that $\mu_{R}$ strongly differs from the local marker $\nu_{R}$, since $[\hat R,\hat P]\!\ne\!0$.

\subsection*{Annihilating the edge-states contribution}

We now introduce two protocols allowing for the annihilation of the undesired edge-states contribution.

The first scheme consists in measuring the rate associated with the dynamical repopulation of the initially-unoccupied bulk bands only, i.e.~disregarding the repopulation of edge states. In practice, this requires the knowledge of the bulk-band structure. Formally, the resulting DIR would probe the quantity $\text{Tr} \, \hat{\mathfrak{C}}$ in Eq.~\eqref{real_space_result}, but with the modified projector operator $\hat Q\!\rightarrow\!\hat Q_{\text{bulk}}$ that excludes the edge states of the spectrum. We have estimated the validity of this approach through a numerical study of the Haldane model with OBC, and found that the topological marker $\nu_{\text{bulk}}\!=\!(1/A_{\text{syst}})\text{Tr } \hat{\mathfrak{C}}_{\text{bulk}}$ resulting from the modification $\hat Q\!\rightarrow\!\hat Q_{\text{bulk}}$ yields the approximate value $\nu_{\text{bulk}}\approx\!-0.85$ for a lattice with $2500$ sites and $\nu_{\text{bulk}}\approx\!-0.91$ for a lattice with $10^4$ sites; these results, which are close to the ideal value $\nu_{\text{LB}}\!=\!-1$, are found to be stable with respect to the Fermi energy (i.e.~to the number of initially populated edge states) and improve as the system size increases. We then validated this scheme through a complete numerical simulation of the full-time-evolution associated with the circularly-driven Haldane model with OBC:~we found that the resulting response coefficient in Eq.~\eqref{main_result} verified $\eta_0\!\approx\!\nu_{\text{bulk}}/\hbar^2$, as estimated from the modified topological marker introduced above. This indicates how restricting the measurement of the depletion rate to the repopulation of bulk states only allows for a satisfactory evaluation of the quantized DIR in Eq.~\eqref{main_result} under realistic conditions.

\begin{figure}[h!]
\resizebox{0.48\textwidth}{!}{\includegraphics*{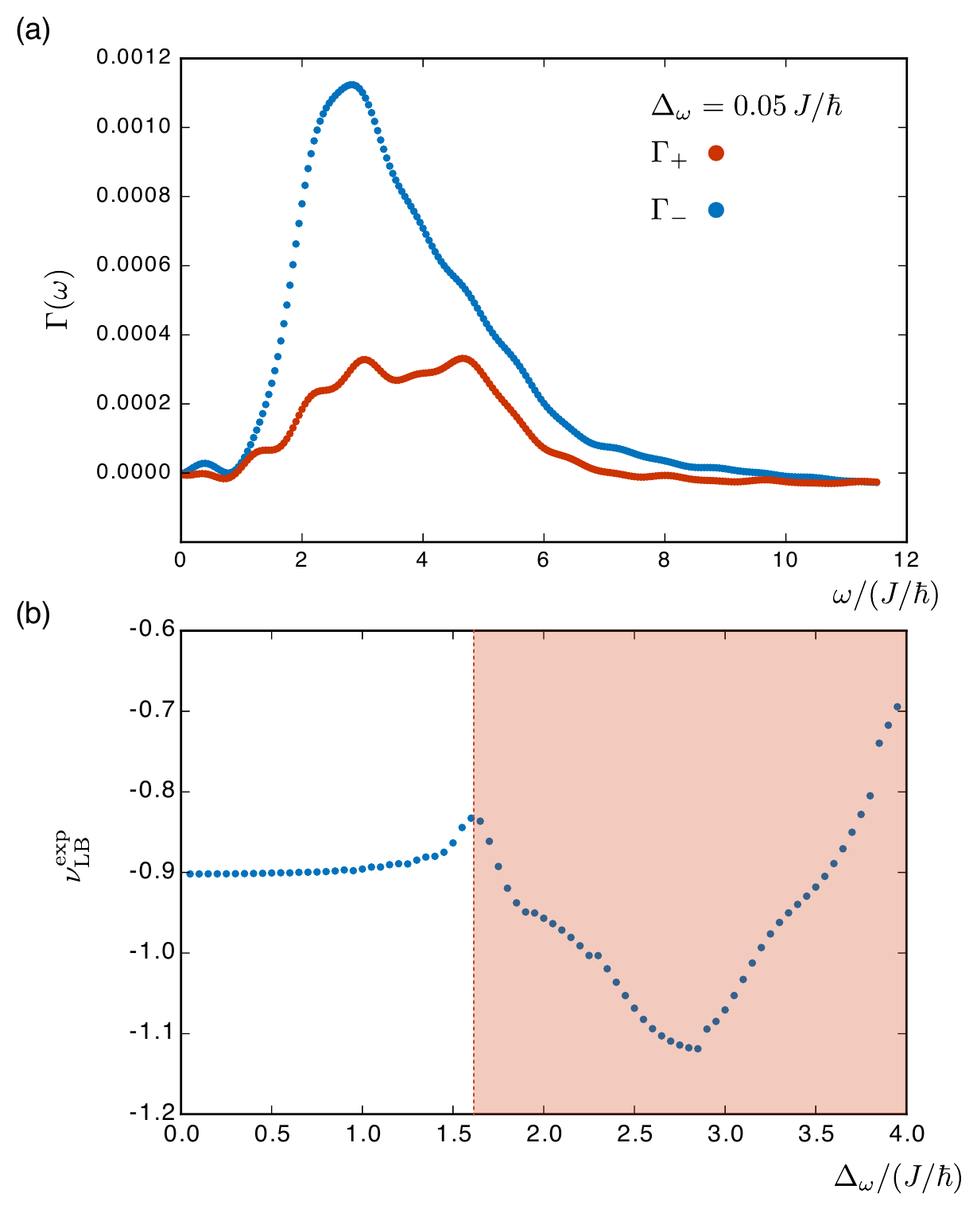}}  
\caption{(a) Depletion rates $\Gamma_{\pm} (\omega)$ extracted from a numerical simulation of the circularly-shaken Haldane model with OBC. The edge-state contribution has been annihilated by initially confining the cloud in a disc of radius $r\!=\!20d$, and then releasing it in a larger lattice of size $120\!\times\!120$, after which the heating protocol (circular drive) was applied; other system parameters are the same as in Fig.~\ref{fig:2}. The rates $\Gamma_{\pm} (\omega_l)$, which are expressed in units of $J/\hbar$, were obtained by measuring the number of excited particles after a time $t\!=\!4\hbar/J$, for fixed values of $\omega_l$ separated by $\Delta_{\omega}\!=\!0.05 J/\hbar$. (b) Approximate value for the Chern number of the populated band $\nu_{\text{LB}}^{\text{exp}}$, as extracted from the numerical rates and Eq.~\eqref{main_result}, and represented as a function of the step $\Delta_{\omega}$ used to sample the drive frequencies; note that the area $A_{\text{syst}}$ entering Eq.~\eqref{main_result} corresponds to the initial area of the cloud in the trap-release protocol. A satisfactory measure is reached when this sampling accurately probes the resonant peaks; we find $\Delta_{\omega}\!\lesssim\!0.5 J/\hbar$ (i.e.~at least 20 different frequencies) for an observation time $t\!=\!4\hbar/J$. The saturation value $\nu_{\text{LB}}^{\text{exp}}\!\approx\!-0.9$ is limited by the fraction of particles populating the upper band, after abruptly removing the confinement, and can be improved by further increasing the initial radius $r$ (or by softening the trap release). 
}
 \label{fig:4} 
\end{figure}

We then explore a more powerful scheme, which does not rely on the knowledge of the bulk band structure. Inspired by Ref.~\cite{Dauphin}, we propose to initially prepare the system in the presence of a tight confining trap, and then to release the latter before performing the heating protocol. In this case, edge states associated with the full (unconfined) lattice remain unpopulated, as they do not couple to the time-evolving cloud upon the drive. We have validated this scheme numerically through a complete  time-evolution simulation of the circularly-driven Haldane model, and we summarize the results in Fig.~\ref{fig:4}. Figure~\ref{fig:4}(a) shows the depletion rates $\Gamma_{\pm} (\omega_l)$, where $\omega_l$ are the many sampled frequencies; here, features of the sinc-squared function are visible due to the finite observation time [Eq.~\eqref{rate_gen_real}]. Figure~\ref{fig:4}(b) shows the value of the extracted Chern number $\nu_{\text{LB}}^{\text{exp}}$, as a function of the frequency sampling step $\Delta_{\omega}$; these values were obtained by comparing Eq.~\eqref{main_result} to  the numerical DIR, $\Delta\Gamma^{\text{int}}\!=\!\sum_l \left [\Gamma_{+} (\omega_l)\!-\!\Gamma_{-} (\omega_l) \right ]\Delta_{\omega}/2$. In this protocol, a residual deviation from the ideal DIR quantization is still visible, even in the limit $\Delta_{\omega}\!\rightarrow\!0$; see the saturation value $\nu_{\text{LB}}^{\text{exp}}\!\approx\!-0.9$ in Fig.~\ref{fig:4}(b).~This is mainly due to the finite population of the higher band upon abruptly releasing the trap; we note that this weak effect is more pronounced for systems for which the Berry curvature is peaked close to the band-gap (e.g.~the Haldane model), and that it can be reduced by either increasing the initial size of the cloud or by softening the release of the trap. This numerical study, based on a simulation of the full-time dynamics in real space, demonstrates the validity and robustness of this trap-release protocole under reasonable experimental conditions, i.e.~an observation time of a few hopping period and a limited number of sampled frequencies $\omega_l$. See also Fig.~S1 in the Supplementary Materials. 

We point out that the numerical results shown in Fig.~\ref{fig:4} were obtained by initially confining the cloud using an infinitely abrupt circular trap, which can indeed be designed in experiments~\cite{box2}. Besides, we stress that similar results would be obtained in more standard setups featuring smooth (harmonic) traps~\cite{Cocks,Goldman_Gerbier,Goldman_edge}; in such configurations, the trap-release protocol would then correspond to a significant change in the trap frequency (see Ref.~\cite{Dauphin}, where bulk topological responses were numerically investigated under such a protocol).

\subsection*{Depletion rates and topology: Beyond 2D lattices}

We now illustrate how differential depletion rates associated with circular drive can probe topological matter in higher dimensions. We discuss two generic but distinct effects, which we concretely illustrate with Weyl-semimetal Hamiltonians~\cite{Hosur:2013eb,Weng2015,Huang2015,SIN15,Lv2015}.

The first effect stems from a direct generalization of the 2D analysis; it relies on noting that the expression for $\Delta\Gamma^{\mathrm{int}}$ given in Eq.~\eqref{general} does not depend on dimensionality, provided we consider spatial dimensions $D\!>\!1$, where a chiral time-modulation is indeed well defined. In general, however, the sum over $\bs k$ in Eq.~\eqref{general} involves a $D-$dimensional Brillouin zone (FBZ), which leads to a non-quantized result. To see this, we generalize the drive operator in Eq.~\eqref{Htime} as $\hat{V}_{\pm}(t)\!=\!E(\hat{a}\pm i \hat{b})e^{i\omega t} \!+\!\mathrm{h.c.},$ where $(\hat{a},\hat{b})$ are position operators defining the polarization plane. Using these notations, and considering the 3D case ($D\!=\!3$), the DIR in Eq.~\eqref{general} can  be written as
\begin{equation}
\label{eq:3DGamma}
 \Delta\Gamma^{\mathrm{int}}/V_{\text{syst}} = \eta_{\text{3D}}E^2, \quad   \eta_{\text{3D}}= \bs K \cdot (\bs{1}_a \times \bs{1}_b)/2\pi \hbar^2 ,
\end{equation}
where $V_{\text{syst}}$ is the volume and $\bs{K}$ is a vector with units of momentum. As could have been anticipated from a 3D generalization of Eq.~\eqref{final_dichroism_rate}, the response coefficient $\eta_{\text{3D}}$ is directly analogous to the general expression for the Hall conductivity in 3D, $\sigma^{ab}\!=\!(e^2/h)\epsilon^{abc}K_c/2\pi$, where $\bs K$ is known to contain information on the topology of the bands~\cite{Haldane04,Haldane14}. For instance, in the simplest case of a stacking of 2D Chern insulators,  piled up along $z$ with  inter-plane separation $d_z$, and which is driven by a circular drive polarized in the $x\!-\!y$ plane, we find the DIR in Eq.~\eqref{eq:3DGamma} with $K_z\!=\!2 \pi \nu_{\text{LB}}/d_z$, where $\nu_{\text{LB}}$ is the Chern number associated with each plane.  In the context of time-reversal-breaking Weyl semimetals, a similar calculation identifies $\bs K\!=\!\nu_{\text{tot}} \bs d_W$, where $\bs{d}_W$ is a vector connecting the Weyl nodes in $k$-space and $\nu_{\text{tot}}$ is the total Chern number of the occupied bands between these nodes~\cite{Burkov:2011de,Zyuzin:2012ca,Grushin:2012cb,Zyuzin2012kl,Goswami:2013jp}. We recall that such Weyl semi-metals can become 3D-QH insulators whenever the Weyl nodes meet at the edge of the FBZ, in which case the DIR in Eq.~\eqref{eq:3DGamma} is characterized by $\bs{K}\!=\!\nu_{\mathrm{tot}}\bs{G}^0$, where $\bs{G}^0$ is a primitive reciprocal lattice vector~\cite{KH93,Haldane14}. These results indicate that, similarly to the Hall conductivity in 3D, the DIR in Eq.~\eqref{eq:3DGamma} can probe topological properties of the bands, as well as non-universal properties (e.g.~the Weyl node separation).

The above considerations require a protocol involving an integration over frequencies [Eq.~\eqref{general}]. Our second protocol only involves a single frequency $\omega$, and ultimately leads to a quantized signature stemming from the FGR in 3D lattices. It is based on a two-band analysis and builds on the observation that the differential current $\Delta j^a\!=\!(j^a_{+}\!-\! j^a_{-})/2$ is directly related to the \emph{local} differential rate $\Delta\Gamma(\bs k; \omega)\!=\![\Gamma_+ (\bs k; \omega)\!-\!\Gamma_- (\bs k; \omega)]/2$, see Eq.~\eqref{rate_gen}, through
\begin{equation}
	\dfrac{d\Delta j^{a}}{dt} = \int_{\text{FBZ}} \frac{d^3k}{(2\pi)^3} (v^a_1-v^a_0)\Delta\Gamma(\bs k;\omega) , \label{delta_current}
\end{equation}
where $v^a_{0,1}\!=\!\partial_{k_{a}}[\varepsilon_{0,1}(\bs k)]$ are the band velocities in the two bands, and where we set the charge $q\!=\!1$. In this protocol, we take the polarization plane of the drive to be perpendicular to the direction $a$. Then, noting that the $\delta$-function in $\Delta\Gamma(\bs k; \omega)$ [Eq.~\eqref{rate_gen}] defines a surface in $k$-space $S$ orthogonal to the gradient $\partial_{k_a} (\varepsilon_{1}-\varepsilon_{0})$, leads to
\begin{equation}
\label{eq:qmonop}
	\sum_{a=x,y,z}\dfrac{d\Delta j^{a}}{dt}  = \frac{E^2}{8\pi^2\hbar^2} \int_S d\mathbf{S}\cdot\boldsymbol{\Omega} = \frac{E^2}{4\pi \hbar^2} \sum_i C_i,
\end{equation}
where $\boldsymbol{\Omega}$ denotes the Berry-curvature vector~\cite{Xiao_review} of the LB, and where the last sum extends over all momentum space monopoles $i$ enclosed by the surface $S$ with integer charge $C_i$. If all monopoles in the FBZ lie inside $S$ then $\sum_i C_i\!=\!0$. However, if $S$ encloses an uneven number of positive and negative monopoles, e.g.~when Weyl nodes of opposite chirality lie at different energies, the quantity in Eq.~\eqref{eq:qmonop} is non-zero and quantized. This result is an instance of the quantized non-linear photogalvanic effect~\cite{Sipe}, predicted in Ref.~\cite{JGM11} for mirror-free Weyl semimetals; this alternative FGR-derivation highlights the deep connection of this quantized phenomenon to the quantization law in Eq.~\eqref{main_result}. In particular, it suggests how such an effect could be observed in a cold-atom realization of Weyl semimetals~\cite{DKL15}:~if the band velocities $v^a_{0,1}$ are known, measuring the local rate $\Delta \Gamma (\bs k;\omega)$ from wave-packets can lead to a quantized measurement through the integration in Eq.~\eqref{delta_current}. Alternatively, one could probe the current~\cite{Catherine_current} or the related center-of-mass velocity~\cite{Dauphin,PZO16,AidelsburgerNat}, giving directly access to the left hand side of Eq.~\eqref{eq:qmonop}. We note that similar Berry-curvature effects have been recently investigated in the context of circular dichroism in nodal-line semimetals~\cite{Ying-Liu}.

\vspace{-0.5cm}
\section*{Discussion}

In this work, we demonstrated that the depletion rate of filled Bloch bands can satisfy a quantization law imposed by topology. This quantized effect positions depletion-rate measurements as a powerful and universal probe for topological order in quantum matter. In this context, we emphasized the crucial necessity to isolate the bulk response from any detrimental effects associated with the edge modes, which, as we argued, can be realized by exploiting the highly-controllable environment and tools offered by ultracold-atom setups.

Here, we illustrated this phenomenon by considering the case of 2D Chern insulators subjected to circular drives. However, we anticipate that other drive protocols could lead to distinct quantized responses in higher spatial dimensions $D\!\ge\!3$, offering the possibility of revealing other topological invariants (e.g.~higher-order Chern numbers~\cite{4D_Zhang,4D_atoms,4D_Spielman})  through depletion-rate measurements. We point out that a circular perturbation [Eq.~\eqref{Htime}] applied to a gapped surface of a 3D topological insulator~\cite{quantized_faraday} could reveal the half-integer QH effect [through Eq.~\eqref{final_dichroism_rate}], an unambiguous manifestation of these 3D topological states~\cite{topological_review,half-integer}.

Moreover, we emphasize that the general result in Eq.~\eqref{main_result} could be generalized to interacting systems, as suggested by the real-space approach [Eq.~\eqref{real_space_result}], but also by the general sum-rule analysis leading to Eq.~\eqref{final_dichroism_rate}, which directly relates the DIR to the Hall conductivity of the probed system (and which does not make any assumption regarding the nature of interactions in the latter). In this sense, the quantized DIR introduced in this work could be exploited to probe the topological order (e.g.~the many-body Chern number~\cite{Niu1985}) of interacting systems, such as fractional Chern insulators~\cite{FCI}. For instance, the DIR could directly reveal the fractional nature of the Hall conductivity, a striking signature of fractional Chern insulators~\cite{FCI}, through Eq.~\eqref{final_dichroism_rate}.

Finally, we note that similar schemes could also probe the chiral edge excitations of topological phases~\cite{Goldman_Gerbier}, as well as the spin chirality of strongly-correlated states, as recently suggested in Ref.~\cite{Kitamura:2018}.\\

\vspace{7cm}

\paragraph*{Acknowledgments} We acknowledge M. Aidelsburger, M. Bukov, A. Celi, J.~Dalibard, M. Dalmonte, F. Grusdt, F. de Juan, P. Massignan, J. E. Moore, T. Morimoto, S. Nascimbene, M. S. Rudner,  S. Stringari, C. Weitenberg and M. Zwierlein for insightful discussions.  P.Z. gratefully acknowledges support and hospitality from the International Solvay Institutes, as Jacques Solvay International Chair in Physics 2015, which initiated the present collaboration.

\paragraph*{Funding information}

The work in Brussels is supported by the ERC Starting Grant TopoCold and by the FRS-FNRS (Belgium). A.~D. is financed by the Cellex-ICFO-MPQ fellowship and acknowledges support from Adv. ERC grant OSYRIS, EU grant QUIC (H2020-FETPROACT-2014 No. 641122), EU STREP EQuaM, MINECO (Severo Ochoa grant SEV-2015-0522 and FOQUS FIS2013-46768), Generalitat de Catalunya (SGR 874), Fundacio Privada Cellex, and CERCA Program/Generalitat de Catalunya. A.~G.~G. was supported by the Marie Curie Programme under EC Grant agreement No. 653846. Work at Innsbruck is supported by ERC Synergy Grant UQUAM, and by SFB FOQUS of the Austrian Science Fund.

\paragraph*{Author Contributions} N.G. and P.Z. devised the initial concepts and theory. Analytical calculations were performed by D.T.T., A.G.G, and N.G. Numerical simulations were implemented by D.T.T. and A.D. The manuscript was written by N.G., with inputs from A.G.G. and P.Z. The project was supervised by N.G.

\paragraph*{Competing Interests} The authors declare no competing interests.

\paragraph*{Data and Materials Availability}

All data needed to evaluate the conclusions in the paper are present in the paper. Additional data related to this paper may be requested from the authors ({\blue *ngoldman@ulb.ac.be}).


\begin{thebibliography}{39}
\bibitem{footnote_position} In a tight-binding model, the position operator should be replaced by $\hat x =\sum_j  x_j \vert w_j\rangle\langle w_j \vert$, where $x_j$ is the position of the $j$th site, and $\vert w_j\rangle$ is the Wannier state defined at this site.
\bibitem{footnote:Rabi} We consider the regime where the observation time $t$ is long enough, such that the rotating-wave approximation applies~\cite{CCT_book}; specifically, $t\!\gg\!\hbar/ \Delta_{\text{gap}}$, where $\Delta_{\text{gap}}$ denotes the bandgap above the LB (which sets the minimal relevant frequency $\omega$). Besides, in order to apply the FGR, the time $t$ is assumed to be small compared to the Rabi period, $t\!\ll\!\hbar/E \vert V_{ge} \vert$, where $V_{ge}$ denotes the matrix elements in Eq.~\eqref{rate_gen_real}. For a given model, this imposes constraints on both the  observation time $t$ and the strength of the drive $E$. Considering the Haldane model, with nearest-neighbor hopping amplitude $J$ and a large bandgap of order $\Delta_{\text{gap}}\!\sim\!J$, we find the reasonable ranges $t\!\sim\!1\!-\!10\hbar/J$ and $E\!\sim\!10^{-2}\!-\!10^{-3}J/d$; see also Ref.~\cite{Goldman_Gerbier} for a discussion on realistic parameters regimes.
\bibitem{footnote_omega_range} In practice, the many frequencies $\omega$ can be chosen in the range $\omega\!\in\![\Delta_{\text{gap}} , W_{\text{tot}}]$, where $W_{\text{tot}}$ is the bandwidth of the entire spectrum. The main result in Eq.~\eqref{main_result} builds on the fact that $\nu_{\text{LB}}\!=\!-\sum_{\text{HB}} \nu_{\text{HB}}$, where $\nu_{\text{HB}}$ are the Chern numbers associated with the higher bands ($\text{HB}$).
\end{thebibliography}
 \end{document}


\title{ {\blue Supplementary Materials for \\ Probing topology by ``heating": Quantized circular dichroism in ultracold atoms}}

\author{D.~T. Tran}
\affiliation{Center for Nonlinear Phenomena and Complex Systems,
Universit\'e Libre de Bruxelles, CP 231, Campus Plaine, B-1050 Brussels, Belgium}

\author{A. Dauphin}
\affiliation{ICFO-Institut de Ciencies Fotoniques, The Barcelona Institute of Science and Technology, 08860 Castelldefels (Barcelona), Spain}

\author{A. G. Grushin}
\affiliation{Department of Physics, University of California, Berkeley, California 94720, USA}
\affiliation{Institut N\'eel, CNRS and Université Grenoble Alpes, F-38042 Grenoble, France}


\author{P. Zoller}
\affiliation{International Solvay Institutes,
Universit\'e Libre de Bruxelles, Campus Plaine, B-1050 Brussels, Belgium}
\affiliation{Institute for Theoretical Physics, University of Innsbruck, A-6020 Innsbruck, Austria}
\affiliation{Institute for Quantum Optics and Quantum Information of the Austrian Academy of Sciences, A-6020 Innsbruck, Austria}

\author{N. Goldman${\blue ^*}$}
\affiliation{Center for Nonlinear Phenomena and Complex Systems,
Universit\'e Libre de Bruxelles, CP 231, Campus Plaine, B-1050 Brussels, Belgium}

\date{\today}

\maketitle

\nopagebreak 

In this Supplementary Material, we provide additional results regarding the numerical analysis of the depletion rates $\Gamma_{\pm} (\omega)$, for the two-band Haldane model with periodic boundary conditions (in the topological phase with Chern number $\nu_{\text{LB}}\!=\!-1$); see the main text for the general definition of the rates.\\ 

First of all, we recall that the depletion rates can be written as (see Eq. 7 in the main text)
\begin{align}
&\Gamma_{\pm} (\omega)=\sum_{\bs k}\Gamma_{\pm} (\bs k; \omega), \qquad \qquad \qquad \qquad \qquad \qquad \qquad \text{(A1)} \notag \\
&\Gamma_{\pm} (\bs k; \omega)\!=\! \frac{2 \pi}{\hbar}\sum_{n > 0} \, \vert \mathcal{V}_{n0}^\pm(\bs k)\vert ^2 \, \delta^{(t)} (\varepsilon_n(\bs k) \!-\! \varepsilon_0 (\bs k) \!-\!  \hbar\omega), \notag \\
&\vert \mathcal{V}_{n0}^\pm(\bs k)\vert ^2=\left (E/\hbar \omega \right )^2 \bigg \vert \bigg \langle n (\bs k) \bigg \vert  \frac{1}{i} \frac{\partial \hat H_0}{\partial k_x}\!\mp\! \frac{\partial \hat H_0}{\partial k_y} \bigg \vert 0 (\bs k)   \bigg \rangle \bigg \vert^2,  \hspace{0.652cm}  \text{(A2)}\notag
\end{align}
where $\delta^{(t)}(\varepsilon)\!=\!(2\hbar/\pi t) \sin^2 (\varepsilon t/2\hbar)/\varepsilon^2\!\longrightarrow\!\delta(\varepsilon)$ in the long-time limit. \\

The transition matrix elements $\mathcal{W}_{\pm}\!=\!(2 \pi/\hbar) \vert \mathcal{V}_{10}^\pm (\bs k) \vert ^2$ were numerically evaluated for the two-band Haldane model ($n\!=\!1$) with periodic boundary conditions; these numerical results were shown in Fig.~2 in the main text as a function of the frequency $\omega\!=\!\left [\varepsilon_1(\bs k)\!-\!\varepsilon_0 (\bs k)\right ]/\hbar$; see the caption of Fig.~2 for details. This numerical data allows one to evaluate the rates $\Gamma_{\pm} (\omega)$ in the infinite-time limit, where $\delta^{(t)}(\varepsilon)\!\longrightarrow\!\delta(\varepsilon)$. Practically, we calculated the averaged matrix elements $\langle \mathcal{W}_{\pm} (\omega) \rangle$, defined within each interval of width $\Delta_{\omega}$ (as represented in Fig.~2), and multiplied this quantity by the density of states: $\langle \Gamma (\omega) \rangle = \langle \mathcal{W}_{\pm} (\omega) \rangle \rho (\omega)$, where $\rho (\omega)\!=\!N_{\text{trans}}(\omega)/\hbar \Delta_{\omega}$ and where $N_{\text{trans}}(\omega)$ counts the number of possible transitions within each interval of width $\Delta_{\omega}$. The resulting rates $\langle \Gamma (\omega) \rangle$ are shown in Fig.~S1(a). From this data, one can then evaluate the DIR, $\Delta \Gamma^{\text{int}}\!\approx\!\sum_l \left [ \langle \Gamma_{+}(\omega_l)\rangle - \langle \Gamma_{-}(\omega_l)\rangle \right ] \Delta_{\omega}/2$, where $\{\omega_l \}$ correspond to the sampled values separated by $\Delta_{\omega}$. We verified that this numerical DIR satisfies
\begin{equation}
\Delta \Gamma^{\text{int}}(\hbar^2/A_{\text{syst}}E^2)\!\approx\!-1.00,\notag
\end{equation}
in perfect agreement with the quantized prediction of Eq.~(1) (main text) and the value of the Chern number of the populated band $\nu_{\text{LB}}\!=\!-1$.\\

In order to evaluate finite-time effects, we also calculated the rates $\Gamma_{\pm} (\omega)$ directly from Eq.~(A1) for finite times $t$; as a technical remark, we substitute $\hbar \omega\!\rightarrow\!\left [\varepsilon_1(\bs k)\!-\!\varepsilon_0 (\bs k)\right ]$ in (A2) to avoid unphysical divergence in the limit $\omega\!\rightarrow\!0$. The numerical results, obtained for $t\!=\!100 \hbar/J$, $t\!=\!10 \hbar/J$ and $t\!=\!4 \hbar/J$, are shown in Fig.~S1(b)-(d), respectively. We observe that the rates $\Gamma_{\pm} (\omega)$ evaluated for the long time $t\!=\!100 \hbar/J$  capture the main features of the infinite-time result shown in Fig.~S1(a). As the duration $t$ is decreased, the curves are smoothened and some signatures of the sinc-squared function $\delta^{(t)}(\omega)$ become visible. In particular, we find that the rates $\Gamma_{\pm} (\omega)$ evaluated for the short time $t\!=\!4 \hbar/J$  capture well the main features of the curves illustrated in Fig.~4(a) in the main text; we remind that the latter were obtained by applying the trap-release protocol in a system with open boundary conditions (and that these simulations were performed in real-space, using the full time-dependent Hamiltonian); we also note that the rates in Fig.~4(a) are smaller by one order of magnitude, due to the smaller system size used in these simulations.

The DIR can be directly evaluated through a numerical integration of the rates shown in Fig.~S1(b)-(d). We obtain
\begin{align}
&\Delta \Gamma^{\text{int}}(\hbar^2/A_{\text{syst}}E^2)\!\approx\!-0.99,  \quad \text{ for } \, \,  t= 100 \hbar/J; \notag \\
&\Delta \Gamma^{\text{int}}(\hbar^2/A_{\text{syst}}E^2)\!\approx\!-0.98,  \quad \text{ for } \, \,  t= 10 \hbar/J; \notag \\
&\Delta \Gamma^{\text{int}}(\hbar^2/A_{\text{syst}}E^2)\!\approx\!-0.96,  \quad \text{ for } \, \,  t= 4 \hbar/J. \notag 
\end{align}
Hence, these finite-time results also show good agreement with the quantized prediction of Eq.~(1) (main text) and the value of the Chern number of the populated band $\nu_{\text{LB}}\!=\!-1$.

\begin{figure*}[h!]
\resizebox{0.8\textwidth}{!}{\includegraphics*{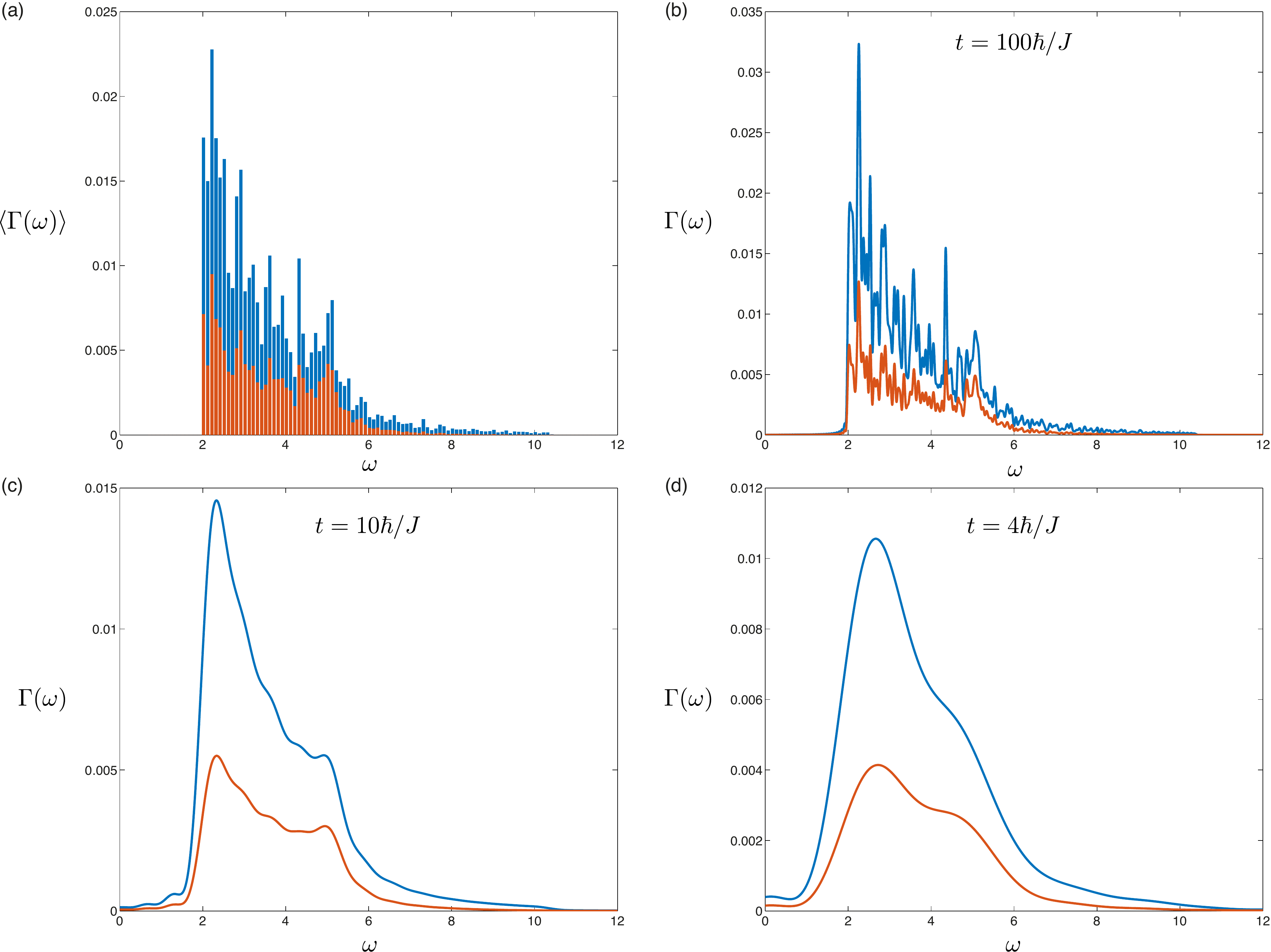}}  
\caption{Depletion rates $\Gamma_{\pm} (\omega)$ as a function of the drive frequency $\omega$ for the driven two-band Haldane model with $10^{4}$ lattice sites and PBC. (a) Rates as calculated from the averaged transition matrix elements $\langle \mathcal{W}_{\pm} (\omega) \rangle$  multiplied by the density of states: $\langle \Gamma_{\pm} (\omega) \rangle = \langle \mathcal{W}_{\pm} (\omega) \rangle \rho (\omega)$. The averages are calculated within each interval of width $\Delta_{\omega}\!=\!0.1J/\hbar$. (b)-(d) Rates as calculated from Eq.~(A1) at finite times $t\!=\!100 \hbar/J$, $t\!=\!10 \hbar/J$ and $t\!=\!4 \hbar/J$, respectively. The system parameters are the same as for Fig.~2 in the main text: the band gap is $\Delta_{\text{gap}}\!\approx\!2J$, where $J$ is the nearest-neighbor hopping amplitude, and the strength of the drive is $E\!=\!0.001 J/d$, where $d$ is the lattice spacing. The rates $\Gamma$ and frequency $\omega$ are expressed in units of $J/\hbar$. As in the main text, the red [resp.~blue] data corresponds to $\Gamma_{+} (\omega)$ [resp.~$\Gamma_{-} (\omega)$].
}
 \label{fig:Sup} 
\end{figure*}